\documentclass[twocolumn,showpacs,preprintnumbers,amsmath,amssymb]{revtex4-2}

\usepackage{graphicx}
\usepackage{dcolumn}
\usepackage{bm}
\usepackage{amsmath}
\usepackage[usenames,dvipsnames,svgnames,table]{xcolor}
\usepackage{enumerate}
\usepackage{mathtools}
\usepackage[colorlinks=true,
            linkcolor=blue,
            urlcolor=blue,
            citecolor=blue]{hyperref}

\newcommand{\be}[1]{\begin{equation}\label{eq:#1}}
\newcommand{\ee}{\end{equation}}
\newcommand{\bea}{\begin{eqnarray}}
\newcommand{\eea}{\end{eqnarray}}

\newcommand{\phd}{\phantom{\dag}}

\newcommand{\up}{^{\phd}}
\newcommand{\noi}{\noindent}
\newcommand{\no}{\nonumber}

\begin{document}

\def\v#1{{\bf #1}}

\title{Controllable critical Josepshon current and $0$-$\pi$ transition in superconductor-insulator-superconductor heterostructures}

\author{G. Livanas$^1$}
\affiliation{$^1$Department of Physics, National Technical University
of Athens, GR-15780 Athens, Greece}

\vskip 1cm
\begin{abstract} 
We investigate Josephson junctions among conventional superconducting wires, in the presence of  externally applied Zeeman fields and supercurrents. We demonstrate that the critical Josephson current and the current-phase relation of the junction depends on the relative orientation of the applied Zeeman fields and supercurrents. The controllability of the Josephson current relies on the stabilization of conventional superconductivity in high magnetic fields, due to the restoration of the opposite momentum opposite spin projection resonance at the Fermi level. We assert that our results can lead to the development of novel superconducting electronics. 
\end{abstract}

\maketitle

The intriguing electromagnetic properties of superconductors have led to the development 
of superconducting electronics \cite{Braginski,Tolpygo} and superconducting spintronics \cite{Linder}, each field related to information processing and storage devices relying on the manipulation of charge and spin degrees of freedom, respectively. Although superconducting devices require cryogenic temperatures, they minimize energy losses and allow for the design of faster and energy-efficient logic and memory units. Basic element of superconducting devices is the Josephson junction (JJ), i. e. two superconducting electrodes connected by a weak link. A dissipationless current, obeying the current-phase relation $I=I_c\sin \phi$, flows in JJs consisting of superconductors locked at macroscopic quantum mechanical phase difference $\phi$ \cite{JosA,JosB}. Besides the phase difference, Josephson junctions are characterised also by the critical current $I_c$, beyond which a voltage drop is developed, i.e. the junction is in the voltage or resistance state, while below this value the charge flow remains dissipationless. 

Controlling the critical Josephson current is the primary objective of superconducting memory devices based on JJs. Establishing two distinct critical current levels in a JJ allows us to store information by switching among the dissipationless and the voltage state. To date, control of the critical Josephson current has been demonstrated in magnetic Josephson junctions (mJJs), i.e. superconductor/ferromagnet/superconductor heterostructures \cite{Bolginov,Lahabi,Li,Parlato,Zhang}. Interplay among the Fraunhofer effect \cite{Schmidt} and the magnetic hysteresis of the ferromagnetic layer, lead to the emergence of two distinct levels for the critical current in these heterostructures. However, the particular junctions demonstrate low normal-state resistance $R_{N}$ and therefore low characteristic voltage $V_c=I_cR_N$  corresponding to high switching time $\tau_s \propto 1/V_c$, rendering them incompatible with rapid flux quantum logic circuits, which are based on superconductor/insulator/superconductor junctions \cite{Kiri}. Attaching additional insulating \cite{Larkin,Ryazanov,Vernik,Karelina} or  ferromagnetic layers \cite{Bell,Baek,Qader,Glick,Satchell} can improve technical characteristics, such as the $\tau_s$, the size and the operating magnetic fields, of the corresponding devices. However, they also increase their complexity.  

An additional degree of freedom of JJs is the phase difference among the SC electrodes. Controllable switches among the 0 and $\pi$ SCs phase difference states can significantly reduce the size and noise in classical \cite{Ustinov,Khabipov} and quantum \cite{Yamashita,Feofanov} computing circuits, respectively. Essentially, a $\pi$-Josephson junction, is a junction where the Josephson current obeys an inverse current-phase relation. Magnetic Josephson junctions can switch among the $0$ and the $\pi$ state based on manipulation of the magnetisation of the ferromagnetic layers \cite{Bethany,Madden,Silaev} or the interfacial spin-orbit coupling \cite{Costa}. Moreover, proposals based on superconductor/normal/superconductor (S/N/S) junctions suggest the use of charge currents \cite{Volkov,Wilh,Basel} and Zeeman fields \cite{Yip} for achieving the $0$-$\pi$ transition in a controllable way. Nevertheless, compatibility with S/I/S junctions hinders the application of mJJs or S/N/S in logic circuits. 

In this article we present a new approach for controlling the critical Josephson current and the $0$-$\pi$ transition in conventional $S/I/S$ Josephson junctions, circumventing in this way the obstacle of compatibility with already developed superconducting devices. Our proposal relies on the combined effect of Zeeman fields and supercurrents in conventional SCs. A conventional SC is an isotropic quantum state formed by coherent pairs of electrons with opposite momenta and opposite spin projection, the singlet Cooper pairs \cite{BCS}. Zeeman fields, i.e. magnetic fields coupled only to the electrons spin, and supercurrents have detrimental effect upon conventional SC. Regarding Zeeman fields, conventional SC survives up to the Chandrasekhar-Clogston limit \cite{Chand,Clog}, $\Delta_0\sqrt{2}$, where  $\Delta_0$ is the superconducting energy gap in the absence of a Zeeman field. Respectively, the Cooper pair velocity $v_s$, related to the dissipationless current flow that a SC wire can sustain, does not exceed $\Delta_0/{\hbar k_F}$, where $k_F$ is the Fermi wavevector \cite{Tinkham}.  

The destructive effect of Zeeman fields and supercurrents on conventional SCs is revealed by examining the energy bands of the system. In the absence of these terms and as long as spatial inversion and spin reversal symmetries are preserved, an opposite momenta, opposite spin projection resonance is present at the Fermi level, favouring the stabilisation of a conventional SC ground state. When a Zeeman field or a current is applied, the particular resonance vanishes. However, their combination can lead to the restoration of the resonance at the Fermi level and therefore allow for the formation of singlet Cooper pairs. Nevertheless, the re-establishment of this resonance occurs only for one of the two possible configurations of the Cooper pairs $k\uparrow, -k\downarrow$ or $k\downarrow, -k\uparrow$, where $k>0$ is the momentum of electrons and $\uparrow,\downarrow$ their two possible spin projections. Therefore, the increase of the current(Zeeman field) in the presence of a Zeeman field(current), leads to a first order transition to a \textit{gapless} conventional SC state with an imbalance among the Cooper pairs of the two configurations. The prevailing configuration of the Cooper pairs can be controlled by the relative orientation of the applied Zeeman fields and supercurrents. We remark that, although a previous work has also demonstrated the persistence of conventional SC in high Zeeman fields beyond the Chandrasekhar-Clogston limit, due to applied voltage bias \cite{Ouassou}, the effect of supercurrents which drives the particular first order transition was not investigated.

\begin{figure}[t!]
\includegraphics[scale=0.5]{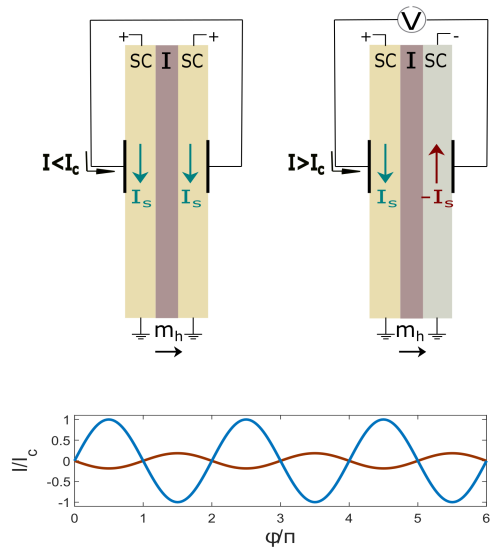} 
\caption{up) Josephson junction among two conventional SC wires in the presence of a Zeeman field $m_h$. Supercurrents $I_s$ are applied in both wires in a parallel (left) and an anti-parallel (right) configuration. When the $I$ perpendicular to the junction is above the critical Josephson current $I_c$, a voltage drop $V$ develops at the edges of the junction. down) Josephson current with respect to phase difference $\phi$ among the two superconductors for the parallel (blue line) and the anti-parallel currents configuration (red line). The two configurations exhibit different critical Josephson current and opposite current-phase relations.}
 \label{fig:1}
\end{figure}

Remarkably, as we present in Fig. \ref{fig:1}, JJs among conventional SCs in the gapless regime exhibit significantly different critical Josephson current, depending on the relative orientation of the applied Zeeman field and supercurrents. Moreover, the current-phase relation is opposite in the two setups. Thus, by applying Zeeman fields and supercurrents in the SCs, we can control both the critical current and the $0$-$\pi$ transition of the JJ. We add, that the Zeeman field can be also applied by depositing the JJ of Fig. \ref{fig:1} on a ferromagnetic insulator \cite{Strambini,Hijano}, enhancing in this way its controllability \cite{Chiba, Avci}. A similar setup has been proposed for S/N/S and S/I/S junctions \cite{Bergeret,Simoni}. We remark that, due to the directionality of the current, the effect of re-emergent SC is prominent in 1D SC wires. Therefore, in the present article we focus on junctions among 1D SC wires. However, the results qualitatively holds also for JJs among quasi-1D and 2D SCs.

Our analysis initiates from the following momentum space Hamiltonian 

\bea
{\cal H} = \sum_{k} \Psi_k^{\dag} \left [ (t\cos J \cos k \right. &-& \mu)\tau_z + t\sin J \sin k  \no \\ &+& \left. m_h\tau_z\sigma_z + \Delta \tau_y\sigma_y \right ] \Psi_k \,, \label{eq:1}
\eea

\noi which incorporates a Zeeman spin-splitting energy  $m_h$ and a current term $t\sin J \sin k$. The spinor $\Psi_k^{\dag} = \left ( \psi_{k,\uparrow}^{\dag},\psi_{k,\downarrow}^{\dag},\psi_{-k,\uparrow}\up,\psi_{-k,\downarrow}\up \right )$, where $\psi_{k,s}^{(\dag)}$ destroys(creates) an electron with momentum $k$ and spin projection $s$, defines the four dimensional Nambu space, while $\mu$ is the chemical potential, $t\cos J \cos k$ is the inversion symmetric kinetic energy of electrons and $\Delta$ the conventional SC field. We employ the Pauli matrices $\tau$ and $\sigma$ acting on particle-hole and spin space respectively. We have introduced the parameter $J$ which corresponds to the gradient of the superconducting order parameter phase, $\Delta e^{-i2Jx}$. Regarding this parameter, SC is sustained as long as $J \cdot \xi_0 <1$, where $\xi_0=\frac{\hbar v_F}{\Delta_0}$ is the coherence length of the SC \cite{Maki}. 

Upon conducting the gauge transformation $\Psi_x \rightarrow e^{-iJx\tau_z}\Psi_x$, where $\Psi_x^{\dag} = \left ( \psi_{x,\uparrow}^{\dag},\psi_{x,\downarrow}^{\dag},\psi_{x,\uparrow}\up,\psi_{x,\downarrow}\up \right )$ is the Nambu spinor in coordinate space and  Fourier transforming to momentum space, we derive the term $t\sin J\sin k$ which relates to the dissipationless current flow $I_s \propto J$ and $t \cos J \cos k$ \cite{LivanasA}.  Nevertheless, the term $t\sin J\sin k$ which connects to the charge current operator $\sum_s\psi_{k,s}^{\dag}\psi_{k,s}\up - \psi_{-k,s}^{\dag}\psi_{-k,s}\up$, can be added irrespective of whether the system is in the superconducting or in the normal phase. When the wire is in the normal phase parameter $J$ associates with a voltage bias applied at the edges of the wire. The eigenenergies of Hamiltonian Eq. \ref{eq:1} take the following form 

\bea
E_{k,s,\pm}= sm_h+t\sin J \sin k \pm \sqrt{\varepsilon_k^2+\Delta^2} \,, \label{eq:2}
\eea

\begin{figure}[t!]
\includegraphics[scale=0.465]{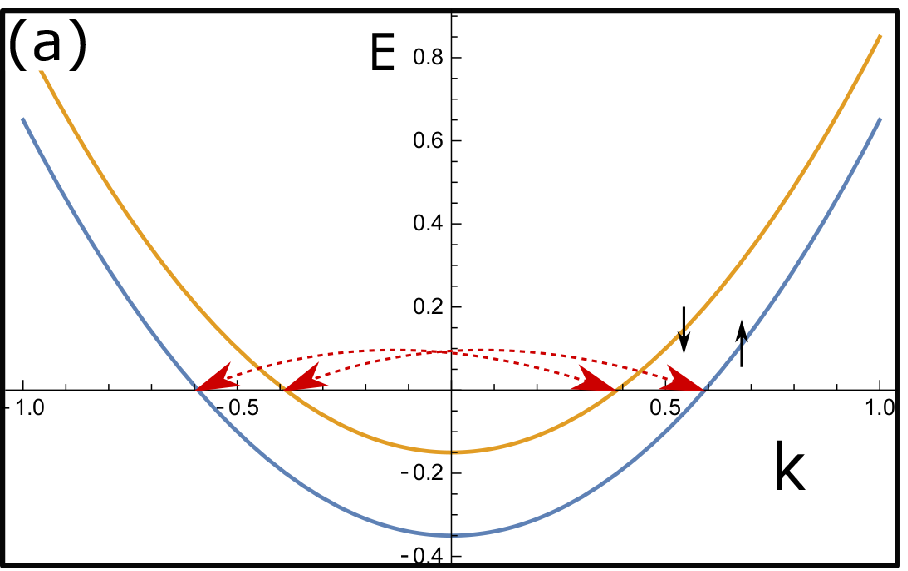} 
\includegraphics[scale=0.465]{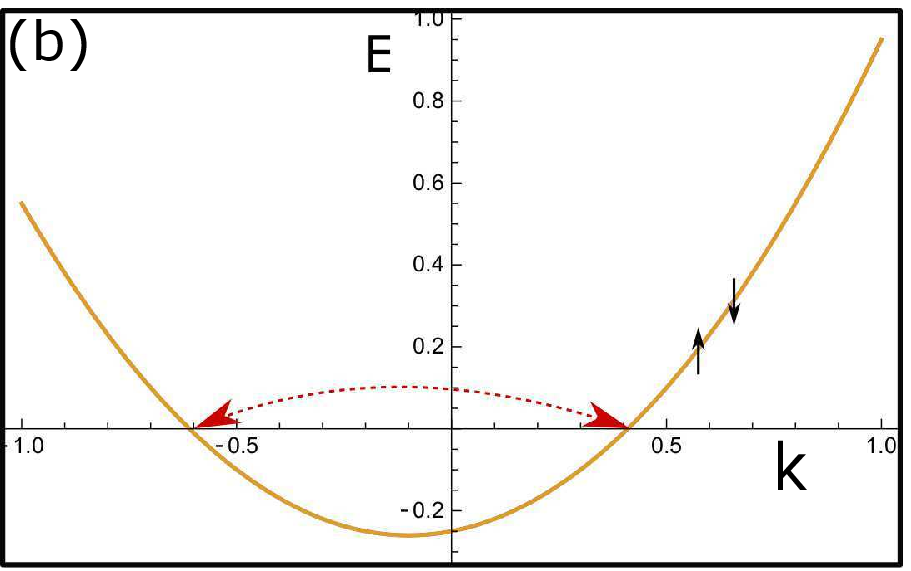} 
\includegraphics[scale=0.465]{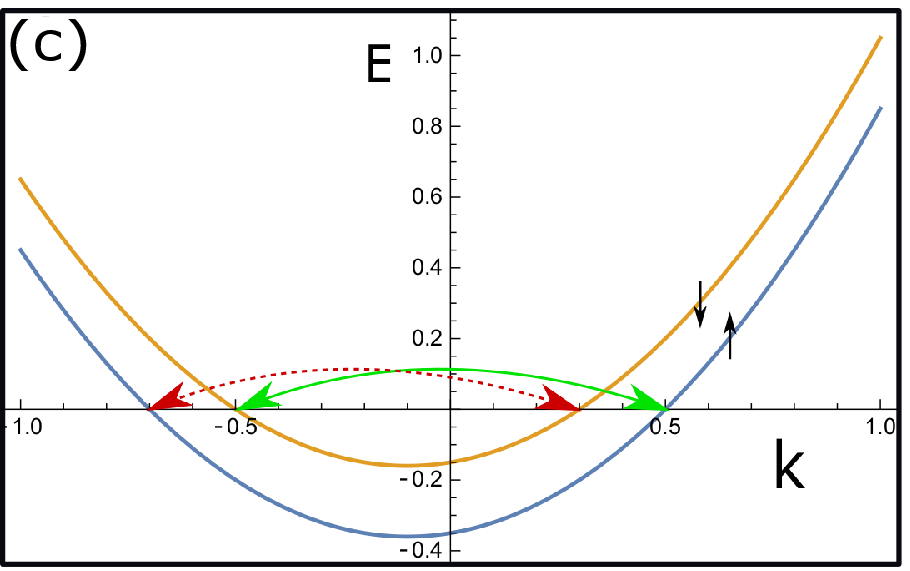} 
\includegraphics[scale=0.465]{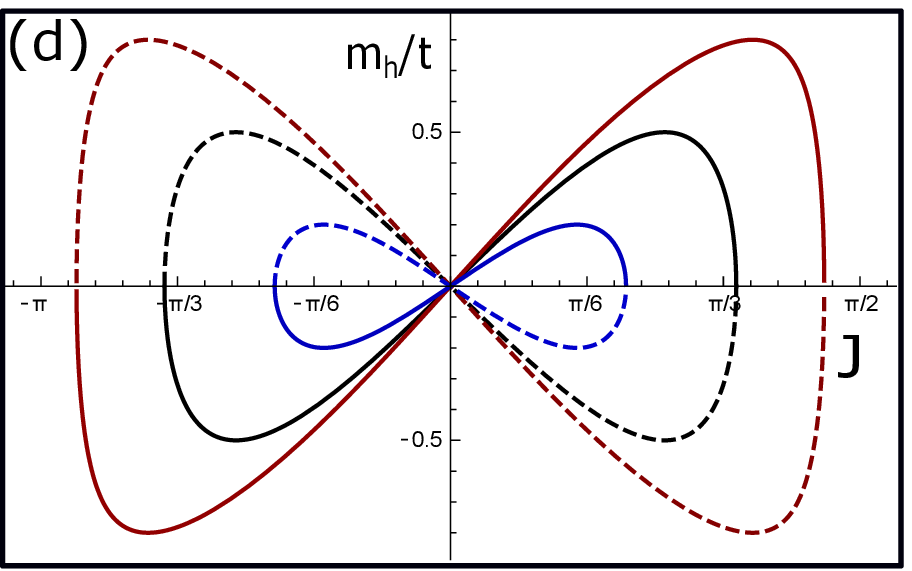} 
\caption{ The energy bands of a conventional SC wire in the presence of a Zeeman field $m_h$ and a supercurrent $J$. (a) For $m_h \neq 0$ and $J=0$ and (b) for $m_h=0$ and $J \neq 0$, the $ks,ks'$ resonance breaks. (c) When both $m_h \neq 0$ and $J \neq 0$, the $ks,-ks'$  resonance can be restored only for the $k\uparrow,-k\downarrow$ or the $k\downarrow,-k\uparrow$ pair, where $k >0$, leading to the re-emergence of conventional SC. (d) The Zeeman field $m_h$ and parameter $J$ for which the resonance is restored for the $k\uparrow,-k\downarrow$ pair (solid line) and the $k\downarrow,-k\uparrow$ pair (dashed line). The chemical potential is $\mu=0.8$ (red line), $\mu=0.5$ (black line) and $\mu=0.2$ (blue line). }
 \label{fig:2}
\end{figure}

\noi with $\varepsilon_k=t\cos J \cos k -\mu$ and $s=+(\uparrow)$ for spin up and $s=-(\downarrow)$ for spin down. In Fig. \ref{fig:2} we present the energy bands of the wire for four distinct cases. We observe that only when both $m_h \neq 0$ and $J \neq 0$, the opposite momentum opposite spin projection resonance can be restored, leading to the re-emergence of conventional SC. However, by examining Eq. \ref{eq:2} for $\Delta=0$, we observe that the resonance is  restored when $sm_h+t\sin J \sin k=0$, only for one of the two configurations $k\uparrow, -k\downarrow$ or $k \downarrow, -k \uparrow$, depending on the relative orientation of the applied Zeeman field $m_h$ and the applied current. In particular, for $ks,-ks'$ configuration  the resonance is restored for $\sin J + \frac{s m_h}{t\sin k_F}=0$, where $k_F = \cos ^{-1} (\frac{\mu}{t \cos J})$ is the Fermi momentum (Fig. \ref{fig:2}(d)). 

\begin{figure}[t!]
\includegraphics[scale=0.19]{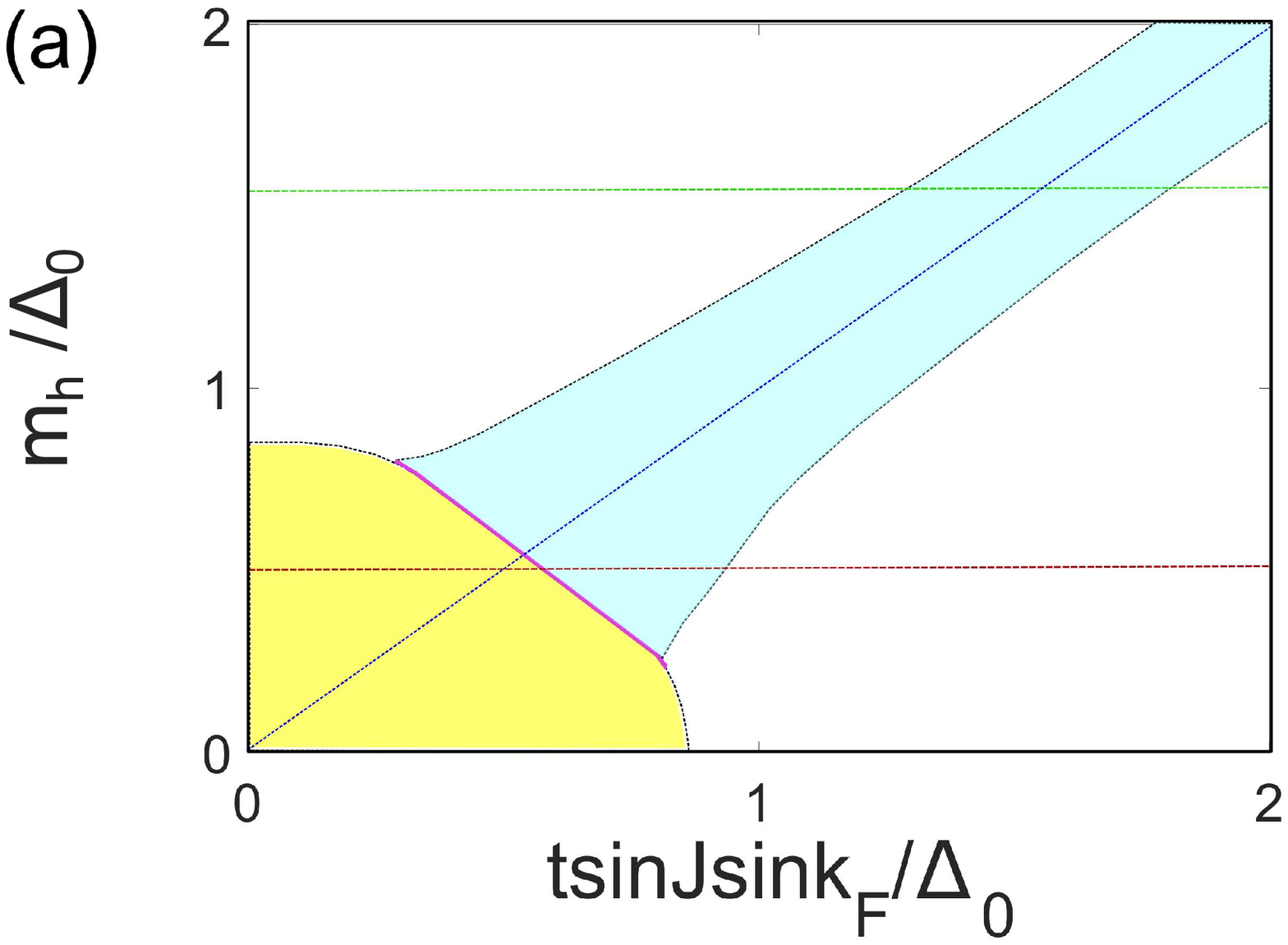} 
\includegraphics[scale=0.19]{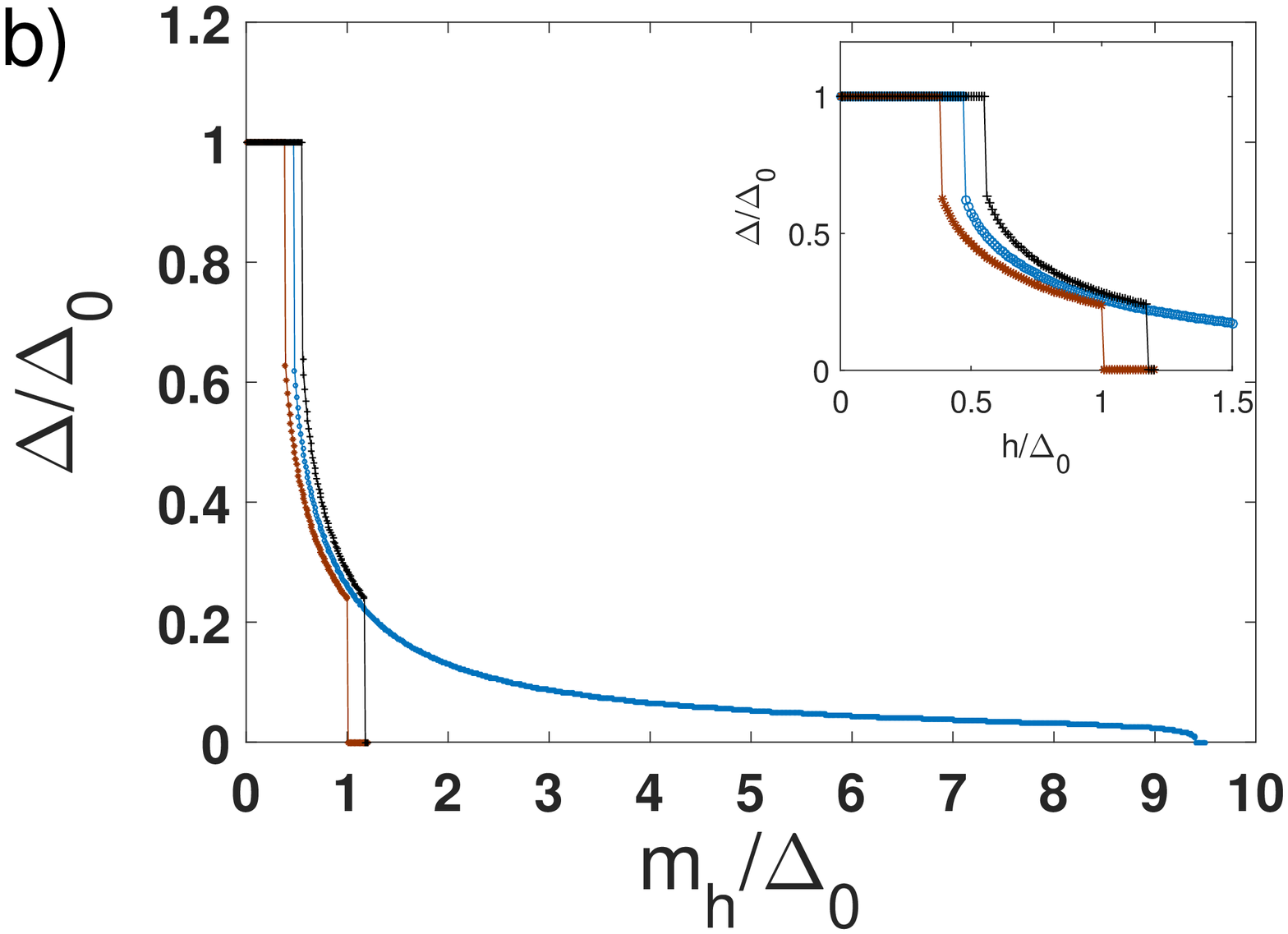} 
\includegraphics[scale=0.19]{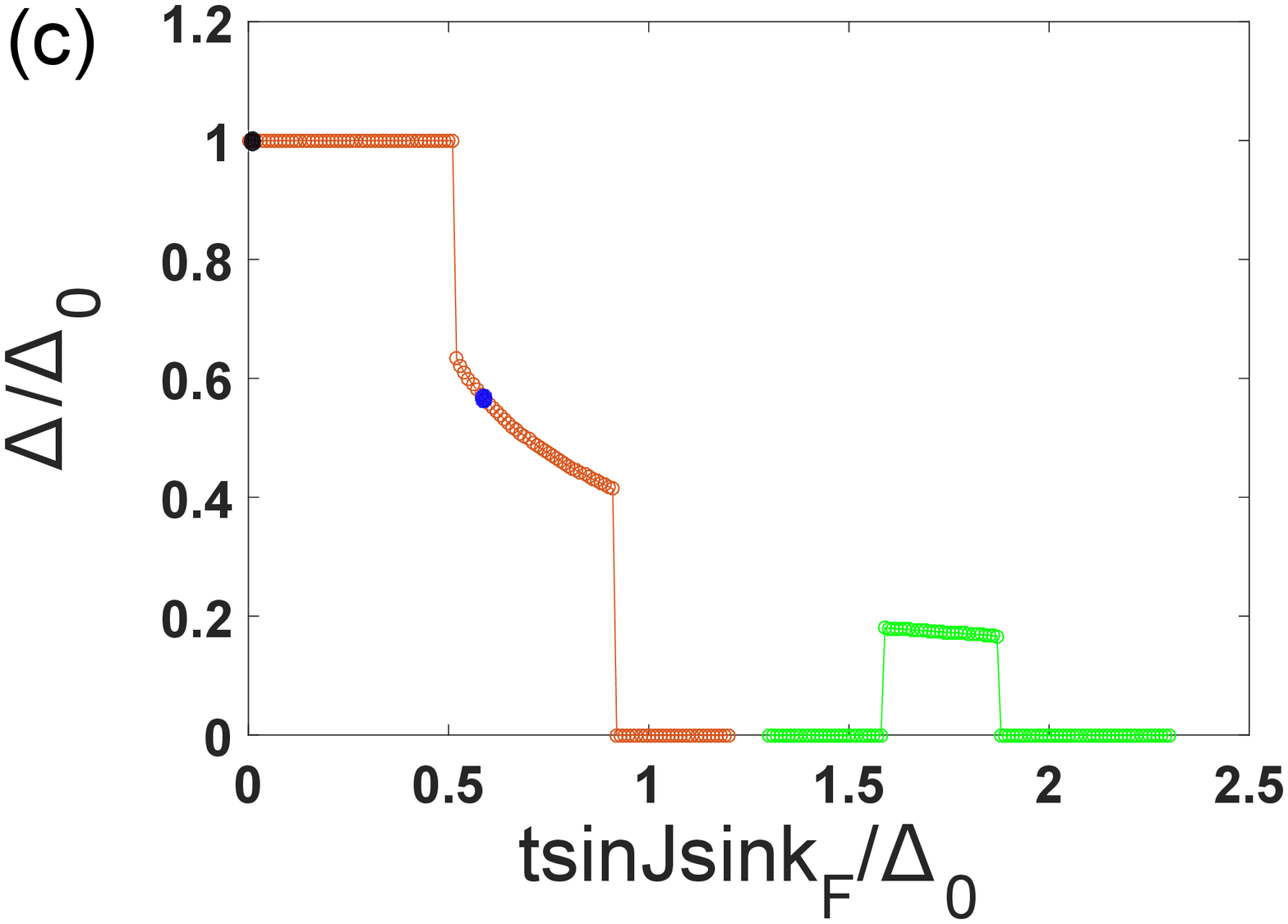} 
\includegraphics[scale=0.19]{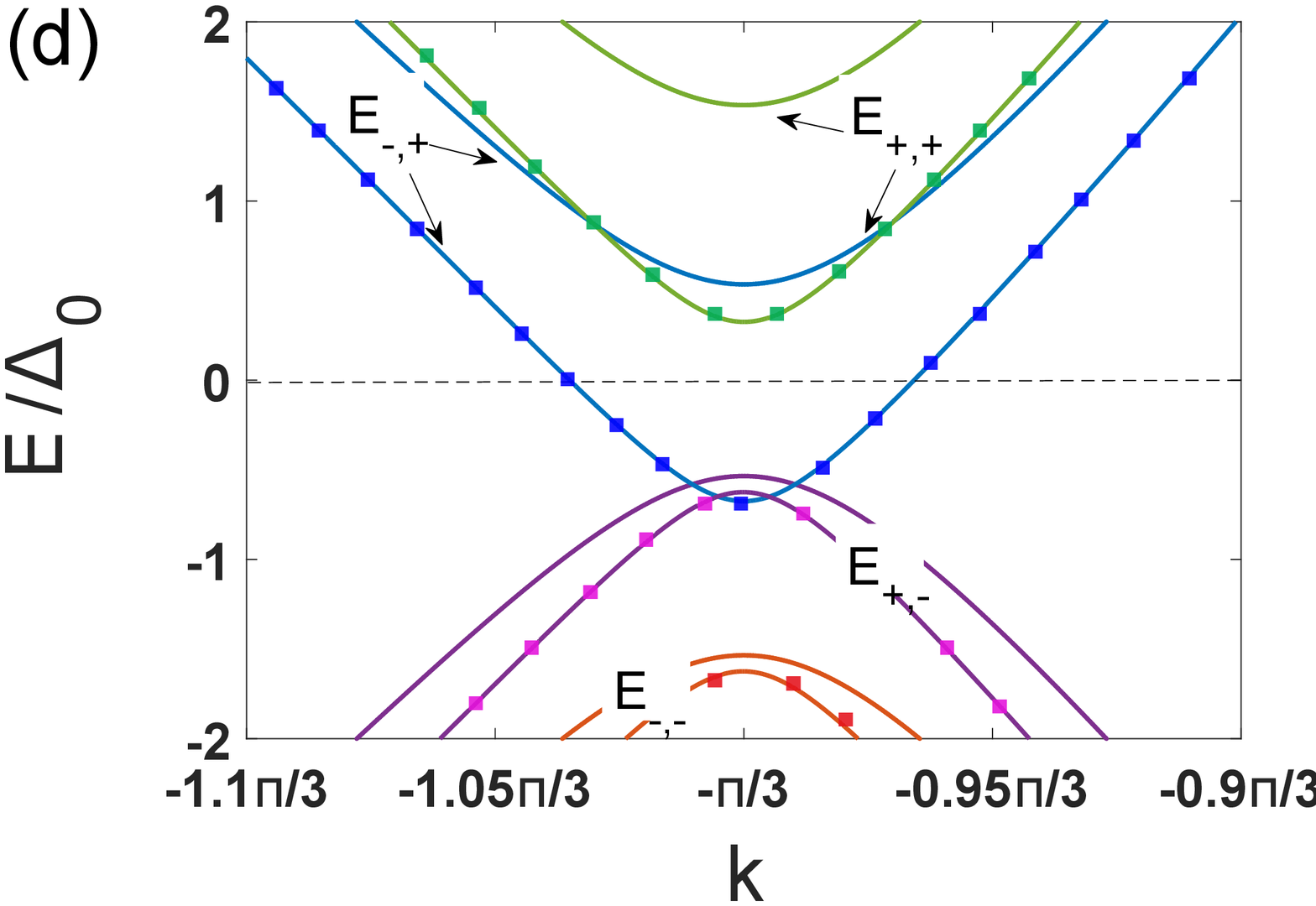} 
\caption{ (a) Phase diagram with respect to the applied Zeeman field $m_h$ and the charge current term $t\sin J \sin k_F$. Purple line denotes a first order transition among fully-gapped SC state with $<\psi_{k,\uparrow}\psi_{-k,\downarrow}>=<\psi_{k,\downarrow}\psi_{-k,\uparrow}>$ (yellow color) and gapless SC state with $<\psi_{k,\uparrow}\psi_{-k,\downarrow}> \neq <\psi_{k,\downarrow}\psi_{-k,\uparrow}>$ (blue color). (b) For   $t\sin J \sin k_F=m_h$, $\Delta$ normalised to the $m_h=J=0$, $\Delta_0$ value, with respect to the applied Zeeman field $m_h$ (blue line, blue line in panel (a)). A first order transition is observed for $m_h=\Delta_0/2$ and a second order transition to the normal state for $m_h>9\Delta_0$. For 
$t\sin J \sin k_F=m_h+0.2 \Delta_0$ (black line) and $t\sin J \sin k_F=m_h-0.2 \Delta_0$ (red line) the transition to the normal state is first order. (c) The normalised SC order parameter with respect to $t\sin J \sin k_F/\Delta_0$ for $m_h=0.5\Delta_0$ (red line and red line in panel (a)) and for $m_h=1.5\Delta$ (green line and green line in panel (a)). (d) For values of $t\sin J \sin k_F$ denoted by the black and blue circle in panel (c) the energy bands of the SC wire (without and with rectangles marks, respectively). We observe that for $t\sin J \sin k_F/\Delta_0=0.65$, $E_{-,+}(E_{+,-})$ crosses the Fermi level for $k<0(k>0)$. }
 \label{fig:3}
\end{figure}

To verify the aforementioned arguments we determine the SC order parameter $\Delta$ which corresponds to the minimum of the free energy of a conventional SC wire in the presence of a Zeeman field and a charge current. The free energy of the wire derives from the following equation

\bea
F=\frac{\Delta^2}{g} - \frac{1}{\beta} \sum_{k,s,\pm} ln(1+e^{-\beta E_{k,s,\pm}}) \,, \label{eq:3}
\eea   

\noi where $g$ is the electrons effective interaction potential and $\beta=k_{B}T$. Results are presented in Fig. \ref{fig:3}, considering a realistic value $\Delta_0=0.01t$ for the SC order parameter. As we observe from the phase diagram, Fig. \ref{fig:3}(a), for finite values of the current, superconductivity is stabilised for Zeeman fields higher than the Chandrasekhar-Clogston limit. In Fig. \ref{fig:3}(b), we present the order parameter with respect to $m_h$, for $t\sin J \sin k_F = m_h$. Notice that $\Delta$ is finite up to $m_h=9\Delta_0$. Moreover, for $m_h=\Delta_0/2$ a first order transition to a SC state with $<\psi_{k,\uparrow}\psi_{-k,\downarrow}> \neq <\psi_{k,\downarrow}\psi_{-k,\uparrow}>$, where $<\psi_{k,s}\psi_{-k,s'}>$ the pairing correlations, i.e. the Cooper pairs, of the $ks,-ks'$ configuration, occurs, while for $m_h=9\Delta_0$ the transition to the normal phase is second order. Nevertheless, as we observe in the same figure the transition to the normal state is first order when $m_h \neq t\sin J \sin k_F$. Finally, for $m_h>\Delta_0/\sqrt{2}$, SC re-emerges, due to the application of a charge current (Fig. 3(c)). Note, that parameter $t\sin J \sin k_F/\Delta_0 \simeq J \xi_0$. We remark, that following this approach we do not examine the stabilization of Fulde-Ferrel-Larkin-Ovchinikov states \cite{Fulde,Ovchi}.

In order to get an insight into phase diagram Fig. \ref{fig:3}(a), we examine the energy spectrum of the SC wire (Fig. \ref{fig:3}(d)). A first order transition occurs when $E_{k,-,+}(E_{k,+,-})$ (for $m_h \cdot J>0$) or $E_{k,+,+}(E_{k,-,-})$ (for $m_h \cdot J<0$) cross the Fermi level for $k<0(k>0)$, i.e. this transition is from a fully-gapped SC state to a gapless SC state.  Another first order transition from the gapless SC to the normal state, occurs when the energy bands for both spin projection cross the Fermi level. As a consequence of the finite energy gap for only one of the two $ks,-ks'$ configurations, an imbalance among the pairing correlations $<\psi_{k,s}\psi_{-k,s'}>$, emerges. Therefore, the particular SC phase can be considered as a mixed $s+p$ state, where $s$ stands for an isotropic s-wave superconducting order parameter and $p$ for an order parameter with $p-$wave symmetry and $S_z=0$ for the spin-z component of the Cooper pairs \cite{LivanasB}. 

In the following, we investigate JJs among conventional SC wires, in the presence of a Zeeman field $m_h$ and a supercurrent. We consider the following Hamiltonian describing the JJ, ${\cal H}= {\cal H}_l^{SC} + {\cal H}_r^{SC}+ {\cal H}_c$, where ${\cal H}_w^{SC} = \sum_{k} \Psi_{w,k}^{\dag}\hat{H}_{k}\Psi_{w,k}$, with

\bea
\hat{H}_{k} = [ \varepsilon_k\tau_z + m_{h,w}\tau_z\sigma_z + t\sin J_w \sin k + \Delta_w\tau_y\sigma_y ] \,, \label{eq:4}
\eea

\noi the Hamiltonian of each SC wire, where index $w=l,r$ denotes the wire. We introduce the extended spinor $\Psi_k^{\dag} = (\Psi_{l,k}^{\dag}, \Psi_{r,k}^{\dag} )$ to model the coupling of the two wires via an insulating layer, by the following expression, ${\cal H}_c= \sum_k\Psi_k^{\dag} t_c\rho_x\tau_z\Psi_k$, where $\bm{\rho}$ are the Pauli matrices acting in the wire bands space. Initiating from Hamiltonian Eq. \ref{eq:4}, we can determine the current flow across the junction based on equation, 

\bea
I \propto \sum_{k,i\omega_n}Tr{\rho_y\hat{G}_{k,i\omega_n}} \,, \label{eq:5}
\eea

\noi where $\hat{G}_{k,i\omega_n}=[i\omega_n-\hat{H}_k]^{-1}$ is the Green's function matrix and $\omega_n = (2n+1)\pi/\beta$ the fermionic Matsubara frequencies. However, to gain insight, we additionally determine the Josephson current of the junction based on equation 
$I= \frac{j}{\beta} \frac{\partial}{\partial \phi}\sum_{i\omega_n}^{k,s} e^{i\phi} F^{l,*}_{ks,-ks',i\omega_n}F^{r}_{ks,-ks',i\omega_n} + c.c.$, where $\phi$ is the phase difference among the two SC wires, $j=-2et_c^2$ and $F$ is the anomalous Green's function \cite{Bruus}. Conducting the summation over $\omega_n$, we derive the following expression for the Josephson current, when $J=m_h=0$ and $T=0$ 

\bea
I_0=2et_c^2\Delta^2\sin \phi \sum_k \frac{1}{E_k^3}\,, \label{eq:6}
\eea
 
\noi where $E_k=\sqrt{\varepsilon_k^2+\Delta^2}$ [see Appendix A]. For $J_l=J_r=\rho$ and $m_{h,r}=m_{h,l}=\rho$, where $\rho>\Delta/2$ the Josephson current reduces to $I_{par} = I_0/2$. This case corresponds to the currents applied in a parallel configuration (Fig. \ref{fig:1}). When the Zeeman field is uniform across the junction, but the currents in the two superconductors are applied in an anti-parallel configuration, $J_l=m_{h,l}=\rho$ and $J_r=-m_{h,r}=\rho$, where $\rho>\Delta/2$, we get the expression

\bea
I_{apar}=-4et_c^2\Delta^2\sin \phi \sum_k \frac{1}{E_k[(2\rho + E_k)^2-E_k^2]}\,. \label{eq:7}
\eea

\begin{figure}[t!]
\includegraphics[scale=0.19]{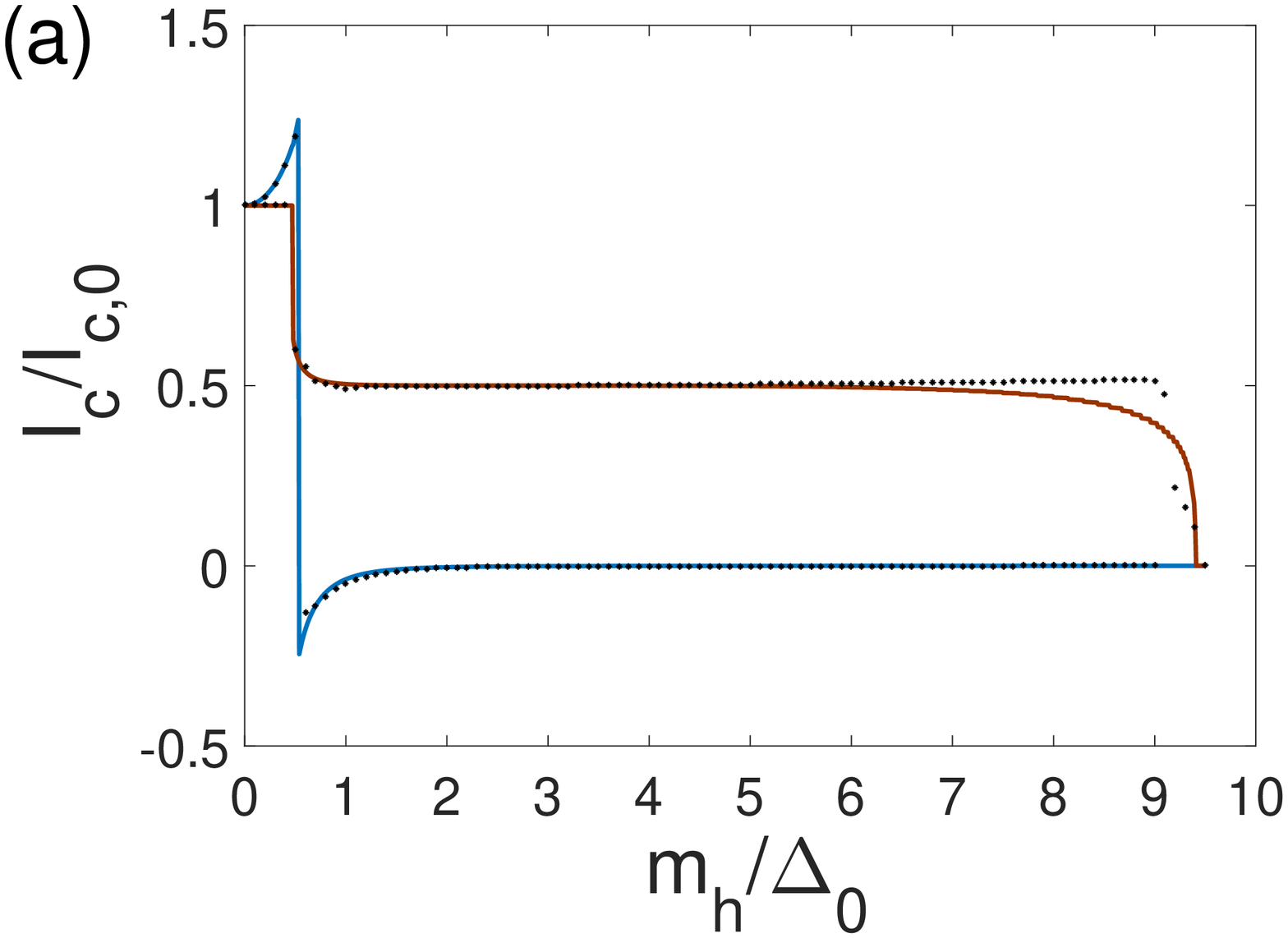} 
\includegraphics[scale=0.19]{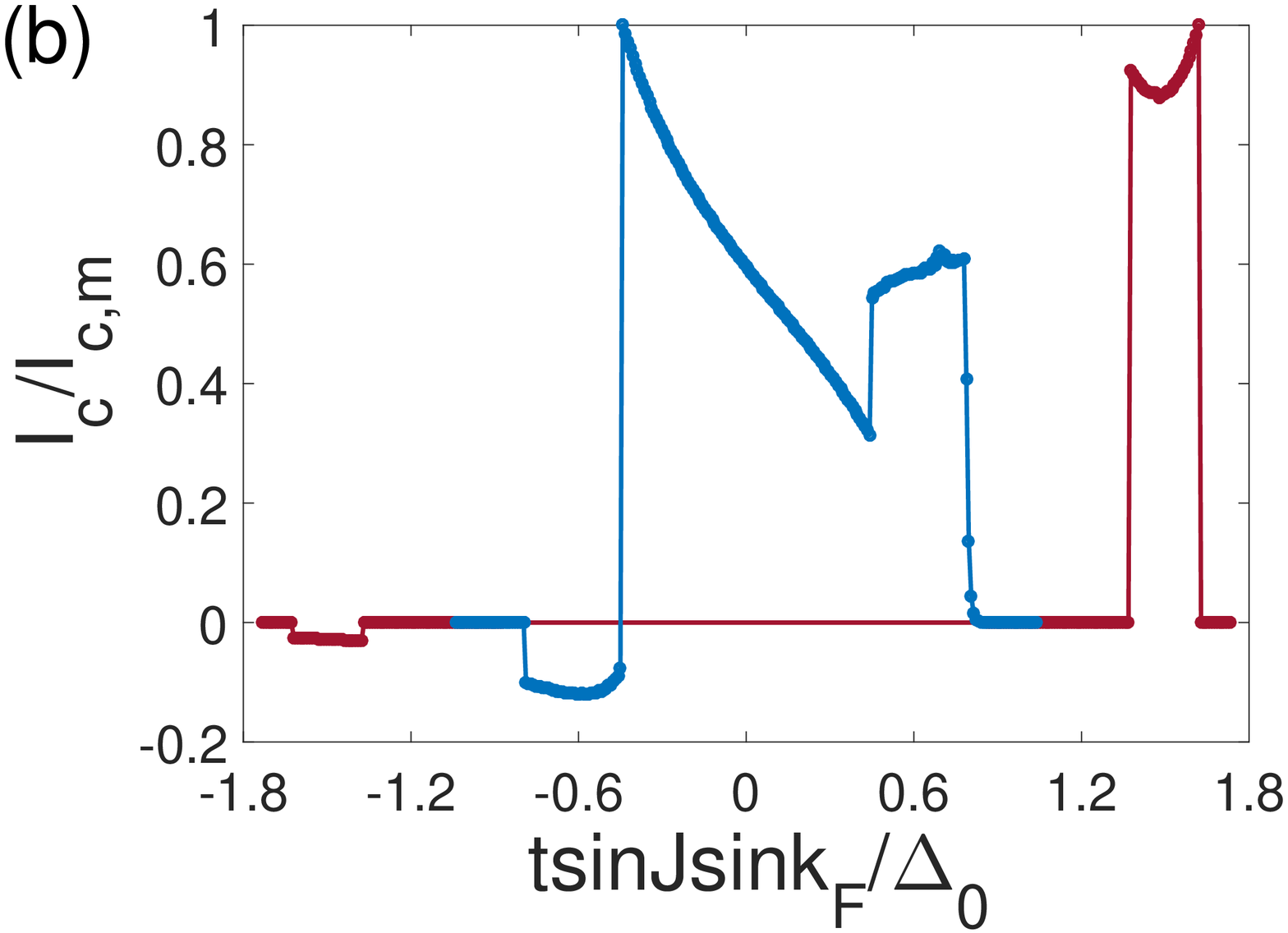}  
\caption{ Critical current for the Josephson junction presented in Fig. \ref{fig:1}. (a) For $m_h=t\sin J \sin k_F$ the critical Josephson current $I_c$ with respect to Zeeman field $m_h$ for the parallel (red line) and the anti-parallel (blue line) configuration. In the parallel configuration, the critical Josephson current reduces to $I_c/2$ beyond $m_h=t\sin J\sin k_F=\Delta_0/2$ and remains constant until the second order transition to the normal state. In the anti-parallel configuration, the critical Josephson current reverse sign for $m_h=\Delta_0/2$ and vanishes for $m_h > 2\Delta_0$. Red and blue lines follow from Eq. \ref{eq:6} and Eq. \ref{eq:7}, respectively, while black marks correspond to results based on Eq. \ref{eq:5}. (b) The critical Josephson current with respect to the applied supercurrent $J_r$ in the right SC wire for $m_h/\Delta_0=0.5$ (blue line) and $m_h/\Delta_0 =1.5$ (red line). The current in the left wire remains constant at $t\sin J_l\sin k_F=0.6\Delta_0$ and $t\sin J_l\sin k_F=1.7\Delta_0$, respectively. We observe that when supercurrent in the second wire reverses direction, the critical Josephson current reverse sign and becomes one (blue line) and two (red line) orders of magnitude smaller that the critical Josephson current in the parallel configuration.}
 \label{fig:4}
\end{figure}

\noi The same expression holds if the currents are applied in the parallel configuration, but the Zeeman fields at the two SCs point at opposite directions. Notice that the Josephson current in the anti-parallel configuration has opposite direction with respect to the current in the parallel configuration. Moreover, $\rho$, which corresponds to the magnitude of the Zeeman field and the current, enters the denominator in the expression for $I_{apar}$, due to the energy difference among states with the same momenta in this case. Since $I_{apar}$ is independent of $\rho$, while $I_{apar}$ reduces rapidly, for large values of the applied Zeeman field and current, $I_{apar}<<I_{par}$. Thus, by manipulating the charge current and/or the Zeeman fields applied on the SCs of the JJ, we can control the flow of the Josephson current. As presented in Fig. \ref{fig:4}(b), corresponding to a JJ where we vary the applied supercurrent on the right SC, for $J_r \simeq J_l$ and $J_r \simeq -J_l$ the critical Josephson current calculated numerically based on Eq. \ref{eq:5} directly from Hamiltonian Eq. \ref{eq:4}, remains relative constant. The magnitude of the Josephson current in these two cases is in qualitatively agreement with Eq. \ref{eq:6} and Eq. \ref{eq:7}. For small values of $J_r$ we observe a linear behaviour of $I_c$, while for $J_r=0$, $I_{c,0} \simeq 1.15I_{par}$ [see Appendix A].

In conclusion, we examined conventional superconductivity in the presence of a Zeeman field and a supercurrent flow. We observe that superconductivity can sustain high magnetic fields beyond the Chandrasekhar-Clogston limit in the presence of a supercurrent. A first order transition to a gapless superconducting state with an imbalance in pairing correlations with opposite spin configuration is observed with increasing supercurrent(Zeeman field) in the presence of a Zeeman field(supercurrent). Depending on the relative orientation of the applied Zeeman fields and supercurrents, conventional JJs exhibit different critical Josephson current and current-phase relations. Since control of the particular junctions can be achieved by means of dissipationless currents, we assert that our results can lead to the development of energy-efficient, multi-functional superconducting devices.

\appendix 
\section{Josephson current}

\noi We introduce the following Hamiltonian describing two conventional SC wires coupled along the transverse to their length direction, in the presence of a Zeeman field and a supercurrent, 

\bea
{\cal H}_{w}=\sum_k\Psi_{w,k}^{\dag}&\Big [& m_{h,w}\tau_z\sigma_z  + [t\cos J_w\cos(k)-\mu]\tau_z \no \\
&+& t\sin J_w\sin(k) + \Delta_w \tau_y\sigma_y \Big] \Psi_{l,k} \,,
\eea
 
\noi where $w=l,r$ denotes the wire and 

\bea
{\cal H}_t=\sum_{k,s}t_c[\psi_{l,k,s}^{\dag}\psi_{r,k,s}\up+\psi_{r,k,s}^{\dag}\psi_{l,k,s}\up]\,, \label{eq:c}
\eea

\noi the coupling part. The Green's function matrix of ${\cal H}_w$, $[i\omega_nI-H_{l,k}]\hat{G}_{l,k}=I$, where $\omega_n=(2n+1)\pi/\beta$ the fermionic Matsubara frequencies. Equating the elements of the matrices in the two part of the above equation we get 

\begin{center}
\begin{tabular}{l}
$[i\omega_n - \epsilon_{k,s}]G_{ks,ks} + s\Delta F_{ks,-ks'}^* = 1$ \no \\ $[i\omega_n + \epsilon_{-k,s}]G_{-ks,-ks} - s\Delta^* F_{ks',-ks} = 1$ \no \\
$[i\omega_n - \epsilon_{k,s}]F_{ks,-ks'} + s\Delta G_{-ks',-ks'} = 0$ \no \\ $[i\omega_n +\epsilon_{-k,s}]F_{ks',-ks}^* - s\Delta^* G_{ks',ks'} = 0$ 
\end{tabular}
\end{center}

\noi where $\epsilon_{k,s}=sm_h+tsJsk+tcJck-\mu$ the energy bands of the wires in the normal phase and $G_{ks,ks'}=-\int_{0}^{\beta}d\tau e^{i\omega_n\tau}<\theta(\tau)\psi_{k,s,\tau}^{\dag}\psi_{k,s',0}\up-\theta(-\tau)\psi_{k,s',0}\up\psi_{k,s,\tau}^{\dag}>$, $F_{ks,-ks'}=-\int_{0}^{\beta}d\tau e^{i\omega_n\tau}<\theta(\tau)\psi_{k,s,\tau}^{\dag}\psi_{-k,s',0}^{\dag}-\theta(-\tau)\psi_{-k,s',0}^{\dag}\psi_{k,s,\tau}^{\dag}>$. Based on the above equations, we determine the elements of the Green's function matrix

\begin{center}
\begin{tabular}{lclcl}
$G_{ks,ks}$&$=$&$\frac{i\omega_n+\varepsilon_{-k,s'}}{(i\omega_n-E_{k,s,+})(i\omega_n-E_{k,s,-})}$ \no \\
\end{tabular}
\end{center}

\begin{center}
\begin{tabular}{lclcl}
$G_{-ks,-ks}$&$=$&$\frac{i\omega_n-\varepsilon_{k,s'}}{(i\omega_n-E_{k,s',+})(i\omega_n-E_{k,s',-})}$  \no \\
\end{tabular}
\end{center}

\begin{center}
\begin{tabular}{lclcl}
$F^*_{ks,-ks'}$&$=$&$\frac{s\Delta^{*}}{(i\omega_n-E_{k,s,+})(i\omega_n-E_{k,s,-})}$ \no \\
\end{tabular}
\end{center}

\noi  where $E_{k,s,\pm}=sm_h + tsJsk \pm \sqrt{\Delta^2+\varepsilon_k^2}$, with $\varepsilon_k=(tcJck-\mu)$, the eigenenergies of the SC wires. Based on Eq. \ref{eq:c} for the coupling of the two wires, the Josephson current emerging, due to a phase difference $\phi$ among the superconductors, derives from equation

\begin{widetext} 
\bea
I&=&-2et_c^2\frac{1}{\beta}\frac{\partial}{\partial \phi}\left [ \sum_{i\omega_n,k,s}F^*_{l,ks,-ks'} F_{r,ks,-ks'}\up  + c.c.  \right] \rightarrow  \no \\
I&=&-2et_c^2\frac{1}{\beta}\frac{\partial}{\partial \phi}\left [ \sum_{i\omega_n,k,s}\frac{\Delta_l}{(i\omega_n - E_{l,k,s,+})(i\omega_n-E_{l,k,s,-})} \frac{\Delta_r^*}{(i\omega_n - E_{r,k,s,+})(i\omega_n - E_{r,k,s,-})} +c.c. \right] \no \\
I&=&et_c^2\Delta_l\Delta_r\sin \phi \left [ \sum_{k,w,s,\pm}\frac{\pm n_F(E_{w,k,s,\pm})}{\sqrt{\Delta_w^2+\varepsilon_k^2}[(s(m_{h,w}-m_{h,w'})+t(sJ_w-sJ_{w'})sk \pm \sqrt{\Delta_w^2+\varepsilon_k^2})^2-(\Delta_{w'}^2+\varepsilon_k^2)]} \right] \label{eq:J}
\eea 
 
\end{widetext}

\noi We examine three distinct cases. First we consider $tsJ_lsk_F=tsJ_rsk_F=m_{h,l}=m_{h,r}=\rho>0$ and $\Delta_l=\Delta_r=\Delta$, i.e. the two superconducting wires are identical. In this case we can drop the $l,r$ index. Thus, Eq. \ref{eq:J} simplifies to

\bea
I&=&4et_c^2\Delta^2\sin \phi\sum_{i\omega_n,k,s} \frac{1}{[(i\omega_n-E_{k,s,+})(i\omega_n-E_{k,s,-})]^2}  \rightarrow \no \\
I&=&-8et_c^2\Delta^2\sin \phi \sum_{k,s}\frac{n(E_{k,s,+})-n(E_{k,s,-})}{(E_{k,\uparrow,+}-E_{k,\uparrow,-})^3} \rightarrow \no \\
I&=&-et_c^2\Delta^2\sin \phi\sum_{k,s}\frac{n(E_{k,s,+})-n(E_{k,s,-})}{\sqrt{\varepsilon_k^2+\Delta^2}^3} \label{eq:para}
\eea

\noi where we have calculated the residue of the second order poles using the formula $Res f(z)_{z=a} =  \left.\partial (z-a)^2f(z)/ \partial z \right |_{z=a}$. Considering $\Delta/t=0.01$ we can simplify momenta summation $\sum_k$, by keeping only terms for which $\varepsilon_k << \Delta $. This leads to replacing $\sum_k \rightarrow d_F c\Delta$, where $d_F$ is the density of states at the Fermi level and $c\Delta$ is the range of energies we  take into account, with $c$ a small parameter.  For T=0, $n(E_{k,s,+})=0$, and for $\rho<\Delta/2$, both spin bands contribute to the current with $n(E_{k,s,-})=1$ and Eq. \ref{eq:para} simplifies to 

\bea
I_{0}&=&2et_c\Delta^2\sin \phi \sum_{k}\frac{1}{\sqrt{\varepsilon_k^2+\Delta^2}^{3}} \no \\
I_{0}&=&2et_c\Delta^2\sin \phi\sum_{k}\frac{1}{E_k^3} \rightarrow \no \\
I_{0}&=&2et_c^2d_F c \sin \phi \,, \label{eq:par0}  
\eea

\noi which is Eq. \ref{eq:6} presented in the main article. However, for $\rho>\Delta/2$, for $k>0$, $n(E_{k,\uparrow,-})=0$ while for $k<0$ $n(E_{k,\downarrow,+})=1$ and cancels the $n(E_{k,\downarrow,-})=1$ term. Therefore, in this case only one spin  band contributes to the Josephson current. Thus, we get 

\bea
I_{par}=et_c^2d_F c\sin \phi=I_0/2 \label{eq:par}
\eea

\noi The second case we examine is where $tsJ_lsk_F=m_{h,l}=\rho>0$ and $tsJ_rsk_F=-m_{h,r}=\rho>0$. In this case we get $E_{l,k,s,\pm }=-E_{r,k,s,\mp}$. For $\rho<\Delta/2$, $E_{w,k,s,-}<0$ while $E_{w,k,s,+}>0$ and both spin bands contribute to the current for $T=0$. Conducting the summation over Matsubara frequencies we get 

\bea
I_{apar}=-4et_c^2\Delta^2\sin \phi \sum_{k,\pm}
\frac{1}{E_{k}[(2\rho \pm E_{k})^2 - E_{k}^2]}\,. \quad 
\eea

\noi Parameter $\rho$ enters the expression for the Josephson current since supercurrents in the SC wires flow in opposite direction. Replacing $\sum_k$ with $d_F c \Delta$,   the above expression  simplifies further to

\bea
I_{apar}&=&-4et_c^2\Delta^2d_Fc\sin \phi \sum_{\pm}
\frac{1}{[(2\rho \pm \Delta)^2-\Delta^2]} \no \\
I_{apar}&=&-2et_c^2\Delta^2d_Fc\sin \phi
\frac{1}{[\rho^2-\Delta^2]} \,. \quad 
\eea

\noi  For $\rho \rightarrow 0$, $I_{apar} \rightarrow I_0$ and as $\rho$ increases $I_{apar}$ increases above $I_0$ quadratically. However, for $\rho>\Delta/2$, $E_{l,k,\uparrow,-}$($E_{r,k,\uparrow,-}$) corresponding to term $\frac{1}{[(2\rho-E_{k})^2-E_{k}^2]}$ for $k>0$($k<0$), becomes positive and does not contribute in the Josephson current. Moreover, $E_{r,k,\downarrow,+}$($E_{l,k,\downarrow,+}$), which corresponds to term $\frac{1}{[(2\rho-E_{k})^2-E_{k}^2]}$ for $k>0$($k<0$), becomes negative and cancels with the corresponding term $E_{l,k,\downarrow,-}$($E_{r,k,\downarrow,-}$) which also corresponds to term $\frac{1}{[(2\rho-E_{k})^2-E_{k}^2]}$. Thus, the above expression is reduced to

\bea
I_{apar}&=&-4et_c^2\Delta^2\sin \phi \sum_{k}
\frac{1}{E_{k}[(2\rho + E_{k})^2 - E_{k}^2]} \no \\
I_{apar}&=&-4et_c^2\Delta^2d_Fc\sin \phi
\frac{1}{[(2\rho + \Delta)^2-\Delta^2]} \no \\
I_{apar}&=&-4et_c^2\Delta^2d_Fc\sin \phi
\frac{1}{2\rho(\rho + \Delta)} \,. \quad 
\eea

\noi which is Eq. \ref{eq:7} presented in the main article. This equation holds at the gapless SC state regime. Notice that, $I_{apar}$ decreases as $\rho$ increases and reverses sign.  \\

\noi The third case we examine is $m_{h,l}=m_{h,r}$ and $tsJ_lsk_F=m_{h,l}=\rho$, where $\rho>\Delta_{l}/2$, but $tsJ_rsk_F=\kappa$ where $\kappa$ is small parameter.  
The eigenenergies of the wires in this case take the form $E_{l,\uparrow(\downarrow),\pm}=+(-)2\rho \pm\sqrt{\Delta_l^2+\varepsilon_k^2}$, $E_{l,\downarrow(\uparrow),\pm}=\pm\sqrt{\Delta_l^2+\varepsilon_k^2}$ for $k>0$($k<0$) and $E_{r,\pm k,s,\pm}=s\rho \pm \kappa \pm\sqrt{\Delta_r^2+\varepsilon_k^2}$. For $T=0$, only $E_{l,\downarrow(\uparrow),-}<0$ for $k>0$($k<0$) and $E_{l,\downarrow,\pm}<0$ for $k<0$ while $E_{r,k,\uparrow,-}<0$  and  $E_{r,k,\downarrow,-}<0$ for any $k$. Therefore, from Eq. \ref{eq:J} we get

\begin{widetext}

\bea
I&=&et_c^2\Delta_l\Delta_r\sin \phi \sum_k\left [\frac{1}{\sqrt{\Delta_l^2+\varepsilon_k^2}[(\rho-\kappa + \sqrt{\Delta_l^2+\varepsilon_k^2})^2-(\Delta_r^2+\varepsilon_k^2)]} \right. \no \\
&+& \left. \frac{1}{\sqrt{\Delta_r^2+\varepsilon_k^2}[(\rho-\kappa-\sqrt{\Delta_r^2+\varepsilon_k^2})^2-(\Delta_l^2+\varepsilon_k^2)]} \right. \no \\
&+& \left. \frac{1}{\sqrt{\Delta_r^2+\varepsilon_k^2}[(\rho-\kappa+ \sqrt{\Delta_r^2+\varepsilon_k^2})^2-(\Delta_l^2+\varepsilon_k^2)]} \right] 
\eea
\end{widetext}

\noi Considering also that  $\Delta_l \simeq \Delta_r/2$ and replacing $\sum_k$ with $d_F c \Delta$ we can further simplify to 

\begin{widetext}
\bea
I=et_c^2d_F c \Delta^2 \sin \phi \left [ \frac{32(\rho-\kappa)}{[9\Delta^2-4(\kappa-\rho)^2][\Delta-2(\kappa-\rho)]} \right ] \label{eq:A11}
\eea
\end{widetext}
 
\noi For $\kappa=0$ and $\rho=0.6\Delta$, we have 

\bea
I=1.15et_c^2d_Fc\sin \phi = 1.15I_{par} 
\eea

\noi while for $\kappa \neq 0$ we get 

\bea
I \simeq I_{par} (1-\frac{\kappa}{\Delta})\,,
\eea

\noi since denominator of Eq. \ref{eq:A11} does not change significantly.








\pagebreak

\end{document}